\documentclass[aoas]{imsart}
\makeatletter
\def\journal@name{}
\makeatother
\usepackage{url}
\usepackage{hyperref}
\RequirePackage{amsthm,amsmath,amsfonts,amssymb}
\usepackage[authoryear]{natbib}
\usepackage{algorithm}
\usepackage{algpseudocode}
\usepackage{graphicx}
\usepackage[textwidth=8em,textsize=small]{todonotes}
\usepackage{xcolor}
\usepackage{comment}
\usepackage{subcaption}
\usepackage[export]{adjustbox}

\startlocaldefs
\theoremstyle{plain}

\newtheorem{theorem}{Theorem}[section]

\newtheorem{proposition}[theorem]{Proposition}
\theoremstyle{remark}
\newtheorem{definition}[theorem]{Definition}

\newtheorem{remark}{Remark}

\endlocaldefs

\begin{document}

\begin{frontmatter}
\title{A Metric Space of Spatial Graphs: Two-Sample Testing, Data Depth, and Application to Cardiac Fibrosis}
\runtitle{Spatial Graphs for Cardiac Fibrosis}

\begin{aug}
\author[A]{\fnms{Anna}~\snm{Calissano}\ead[label=e1]{a.calissano@ucl.ac.uk}},
\author[C,B]{\fnms{Arstanbek}~\snm{Okenov}\ead[label=e2]{a.okenov@lumc.nl}}
\author[C]{\fnms{Katja}~\snm{Zeppenfeld}\ead[label=e4]{k.zeppenfeld@lumc.nl}}
\and
\author[B]{\fnms{Alexander}~\snm{Panfilov}\ead[label=e3]{alexander.panfilov@UGent.be}}
\address[A]{University College London\printead[presep={,\ }]{e1}}
\address[C]{Leiden University Medical Center\printead[presep={,\ }]{e2,e4}}
\address[B]{University of Gent\printead[presep={,\ }]{e3}}
\end{aug}

\begin{abstract}

Cardiac fibrosis — the pathological accumulation of fibroblasts and extracellular matrix in the heart muscle — reduces electrical conductivity and is a leading cause of arrhythmia. Arrhythmic waves typically rotate around non-conducting fibrotic patches, so the geometry and topology of these patches — spatially isolated regions of fibrotic tissue within the heart muscle — play an important role in arrhythmia dynamics. Despite their clinical relevance, these structures remain poorly understood. We address this open problem using histopathological images of human hearts affected by cardiac fibrosis, from which we identified and extracted approximately six million such patches. Each patch was represented as a spatial graph via skeletonization, where nodes are embedded as points in Euclidean space and edges encode geometric properties of the underlying tissue. The core methodological contribution of this work is the introduction of spatial graph space, a metric space equipped with a rotation-invariant Fused Gromov-Wasserstein metric that enables comparison of spatial graphs with differing numbers of nodes and edges. Building on this, we perform a distribution-level statistical testing and depth measures for spatial graphs. To enable the interpretation of the spatial graph sample distribution, we introduce DepthPlot, a novel visualization tool for depth measures in metric spaces.
Applying our methodology to compare patients and the spatial position of patches within the ventricles, we find that fibrotic textures exhibit strong patient-specific features, while some hearts display notable geometric similarities — potentially reflecting shared pathological mutations or other unknown factors. Through quantitative depth measures, we characterize test outcomes via central and peripheral spatial graphs, demonstrating that the proposed framework yields statistically and clinically meaningful insights into fibrotic texture characterization.
\end{abstract}


\end{frontmatter}

\section{Introduction}

\begin{figure}[htbp!]
    \centering
    \includegraphics[width=0.8\linewidth]{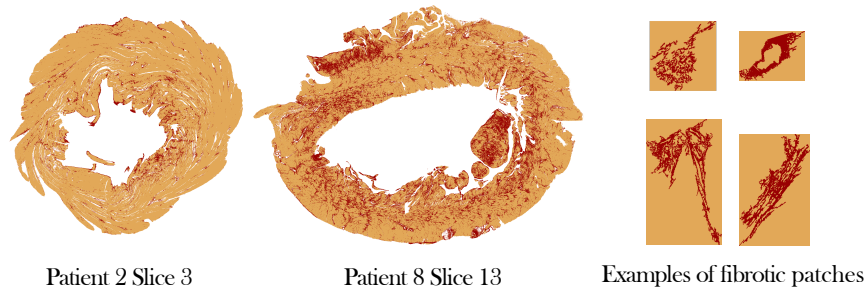}
    \caption{Example of two slices obtained by cutting horizontally the left ventricle belonged to two patients (P2 and P8). The preprocessed images are binary: in orange the heart muscle and in red the fibrotic tissue. On the right, a set of isolated patches is showcased.}
    \label{fig:fibrotic_patches}
\end{figure}

Cardiac arrhythmias are a major public health problem associated with substantial morbidity and mortality, including sudden cardiac death \citep{srinivasan2018sudden}. One of their key precursors is pathological myocardial fibrosis: the excessive accumulation of extracellular matrix proteins, mainly collagen \citep{frangogiannis2021cardiac, de2011fibrosis}. Myocardial fibrosis reduces the reliability of electrical wave propagation through the heart, promoting wave breaks and rotational activity around non-conducting tissue. The probability of such events depends not only on the amount of fibrosis but also on the geometric features of the fibrotic textures \citep{alonso2013reentry, balaban2018fibrosis, ten2007influence}. Despite its clinical relevance, detailed characterization of fibrotic texture geometry remains elusive. Obtaining it requires tedious histological studies that cannot be performed in vivo, while clinical imaging methods cannot resolve fine fibrotic structure (e.g. LGE-Cardiovascular Magnetic Resonance). As a result, existing computational models rely on generic representations of fibrotic textures \citep{nezlobinsky2020anisotropic, spach2000effects, jacquemet2007modelling, kudryashova2019self, boyle2019computationally}, and only qualitative visual classifications exist in clinical settings \citep{de2011fibrosis}. Understanding the true geometry of fibrotic textures could substantially improve modeling approaches and potentially link texture features to specific underlying cardiac conditions. In this paper, we address this open question using a comprehensive histological dataset from non-ischemic cardiomyopathy patients \citep{glashan2018whole}. Such patients are characterized by complex fibrotic textures with multiple non-conducting scars that strongly promote arrhythmias. Since arrhythmias predominantly arise from electrical waves rotating around such scars, we focus on the geometric and statistical characterization of isolated non-conducting regions of fibrotic tissue extracted from the histological images, which we hereafter refer to as patches (see Figure \ref{fig:fibrotic_patches} for examples).

\subsection{Statistical Challenges}
Analyzing a set of images of different shapes, with different topological characteristics, and with high boundary complexity is not straightforward. The field of shape analysis has seen significant growth over the past two decades (see \cite{marron2021object} for a latest overview). This growth is driven by the availability of high-dimensional data that captures objects of interest across various contexts and applications. Starting from the hypothesis that the shape of an object carries information about the function, the category and the role of the object itself, numerous methods have been developed for comparing, summarizing, modeling, testing, and tracking shapes. Statistical shape analysis relies on mathematical representations of a the shape and the definition of an appropriate metrics. Two main representations have been proposed: discrete (e.g., \cite{kendall1984shape,small2012statistical,dryden2016statistical}), and continuous (e.g., \cite{younes1998computable, srivastava2016functional, jermyn2017elastic}). Different representations come with different metric to measure the distance or dissimilarities between the shapes: from elastic metric \citep{srivastava2016functional}, to kernel based metric \citep{vaillant2005surface} and Gromov-Wasserstein distance between objects \citep{memoli2011gromov}, among others. The data studied in this paper show differences in terms of topology, shape, and size; they have highly irregular profiles and high number of holes, resulting in a reticular structure. Thus, the application of standard shape analysis techniques fails for such data type. Due to their branching and porous structures, we decided to study the fibrotic tissue as spatial graphs. Such modeling choice is supported by the intention of capturing both the spatial characteristics and the topology.\\

Spatial graphs are not a novel concept and are called differently in different context. They appear in transport analysis, social network analysis, epidemiology, under the name of spatial networks \citep{doytsher2010querying,barthelemy2011spatial,gou2021understanding}. In these applied contexts, the focus is often on the analysis of one single spatial network. Spatial graphs are also related to geometric graphs, which are graphs with nodes embedded in the Euclidean space and possibly intersecting straight lines as edges \citep{cerny2005geometric}. If the edges are allowed to be continuous curves, the graph is called a topological graph or metric graph \citep{pach1997graphs}. Geometric graphs, topological graph, and metric graph are broadly studied in graph theory, addressing questions about the possible configurations a graph can take given a certain number of nodes and constrains over the edges - see \cite{pach2013beginnings,pach2004geometric} for an historical and technical overview. Last but not least, spatial graphs can be also seen as a simpler version of shape or elastic graphs \citep{guo2020representations}. While in shape graphs the edges encodes an actual shape, in the spatial graphs studied in this paper a linear approximation of the edges' shapes is considered. Although spatial graphs are well-established mathematical objects, their use for the characterization of porous and ramified structures remains largely unexplored. Furthermore, existing embeddings and statistical methodologies for graph-valued data have focused predominantly on abstract graphs, ignoring the spatial component (e.g \cite{ginestet2017hypothesis,kolaczyk2020averages,severn_manifold_2022,calissano_graph_2022,calissano_populations_2024}). We address this gap by developing a statistical framework for the comparison of spatial graphs with heterogeneous numbers of nodes and edges.\\

This work contributes on both the methodological and applied fronts. On the application side, we leverage a unique histological dataset of human hearts from patients with a severe arrhythmogenic phenotype to characterize the geometry and topology of cardiac fibrotic tissue via spatial graphs. On the methodological side, we introduce a metric framework for spatial graphs based on a rotation-invariant Fused Gromov-Wasserstein metric, applicable beyond the cardiac setting to any collection of spatial graph objects. Within this framework, we adapt Anderson's ANOVA as a two sample testing procedure for spatial graph distributions and develop depth measures as an interpretive tool, introducing DepthPlot — a novel adaptation of the boxplot to object data — for visualization of sample distributions in metric spaces. The paper is organized as follows. Section \ref{sec:data} describes the histological dataset. Section \ref{sec:spatial_graphs} introduces the spatial graph representation and the Fused Gromov-Wasserstein metric between spatial graphs. Section \ref{sec:metric_statistics} develops the metric space properties, the testing procedure, depth measures, and the DepthPlot. Section \ref{sec:results} presents results on the modeling of fibrotic tissue as spatial graphs, cross-patient testing, and the empirical distribution of fibrotic patches within the heart via horizontal and vertical exploration. Section \ref{sec:comparison} benchmarks the spatial graph approach against a simpler feature-based representation of the dataset, assessing the added value of the proposed framework.

\section{Dataset}
\label{sec:data}

Built by \cite{glashan2018whole}, the dataset includes $10$ patients with a non-ischemic cardiomyopathy and sustained ventricular arrhythmias, diagnosed prior to death or heart transplantation. Each heart was cut into horizontal slices with a $5$ millimeter distance as visualized in Figure \ref{fig:fibrotic_patches}, producing between $7$ and $19$ slices per patient. Once the slices are obtained, staining was performed using Picrosirius Red, which is a contrast fluid staining collagen in red and the myocardium in yellow. After staining, the fibrotic patches are marked using a 5-point connectivity stencil that produces a set of isolated fibrotic patches (for a total of over 6 million patches). Such patches have high variability in terms of area and shape, as shown in the right panel of Figure \ref{fig:fibrotic_patches}. The analysis focuses on the dataset of bi-dimensional and isolated fibrotic patches obtained from the segmentation of the different slices.

\subsection{Data Representation}
\begin{figure}[htbp!]
    \centering
    \includegraphics[width=0.9\linewidth]{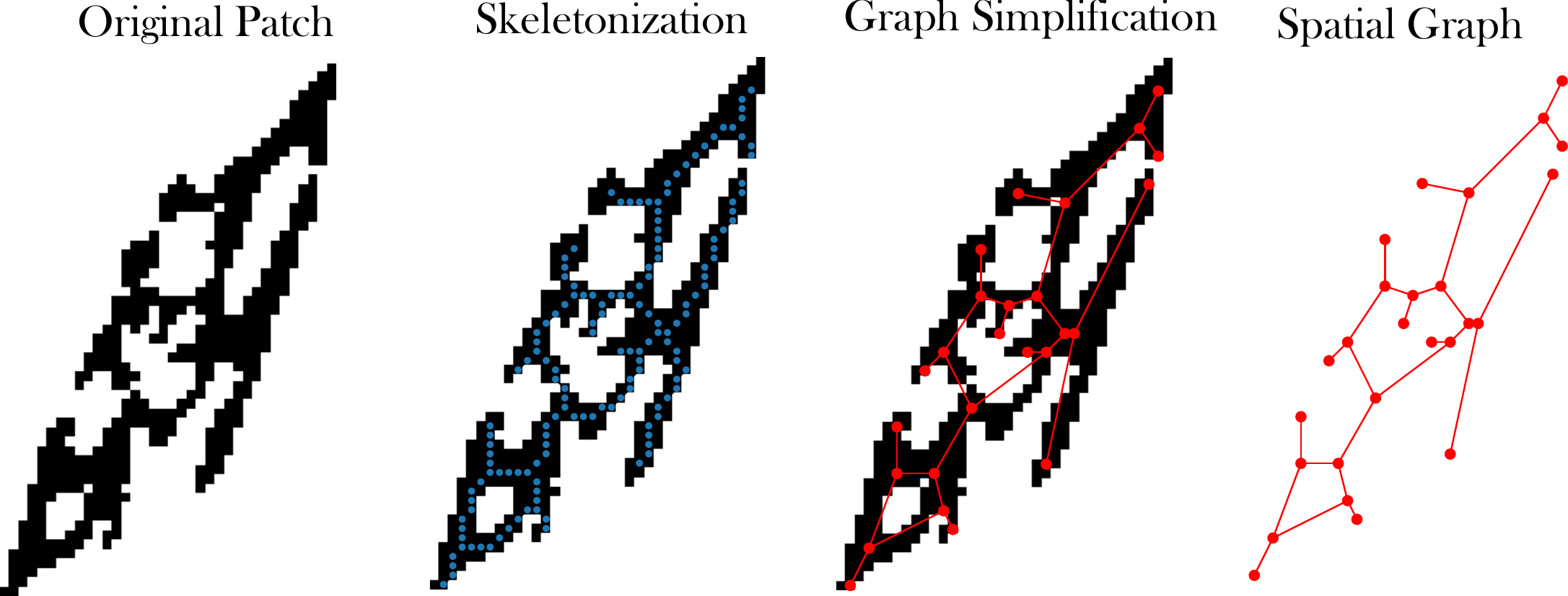}
    \caption{Example of skeletonization procedure of a fibrotic patch, graph simplification via node coarsening, and the resulting spatial graph.}
    \label{fig:skeletonization}
\end{figure}

Consider an image of an isolated fibrotic patch $I\in \mathbb{R}^{n\times m}$ (e.g. the right panel in Figure \ref{fig:fibrotic_patches}). Each image is a binary image where $1$ represents the presence of a fibrotic tissue and $0$ the myocardium or the background. The complete dataset consists in a set of over 6 million 
images of fibrotic tissues. As stated in the introduction, the shapes are too irregular to be analyzed with tools from both continuous and discrete shape analysis representations, so we opted for a graph representation. The graph representation allows for the accounting of the porous structure of the fibrotic tissue (encoded as cycles in the graph representation), the fragmented perimeter (encoded as small branching edges), and the spatial structure (encoded as spatial coordinates of the nodes). As a first step, each image is skeletonized into a structure object, called spatial graph (see Figure \ref{fig:skeletonization}). The literature about extracting a skeleton from an image is rich and we opt for a state of the art method based on thinning \citep{zhang1984fast}.

\section{Spatial Graphs}
\label{sec:spatial_graphs}
A spatial graph is here defined as a product space between two metric measure spaces. We start by defining the different components one by one.
\begin{definition}[Metric Graph]
A metric graph $(X,d_X)$ is a discrete metric space where $X=\{1,\dots, N\}$ is a set of nodes and $d_X:X\times X \rightarrow \mathbb{R}$ is a metric over the set of nodes.
\end{definition}
Definition of metric graphs appears in many different works (e.g. \cite{chowdhury2019gromov,memoli2011gromov}). In this work, the metric $d_X$ is the weighted geodesic distance between the nodes of the graph. We now want to encode the spatial component by adding $A\in\mathbb{R}^{N\times p}$, the spatial coordinates of the nodes.
\begin{definition}[Spatial Graph]
A spatial graph $G$  is the triplet $G=(X\times A, d_X, \mu)$ where\\
    \begin{itemize}
        \item[-] $(X,d_X)$ is a metric graph; 
        \item[-] $A$ is the feature space representing the spatial coordinates of the nodes;
        \item[-] $\mu$ is a fully supported probability measure over $X \times A$.
    \end{itemize}
\end{definition}
The measure considered in this discrete case can be expressed as $\mu=\sum_{i=1}^N h_i \delta_{x_i,a_i}$ where $h\in \Sigma_N$ is an histogram encoding discrete weights over the nodes $\Sigma_N=\{h\in(\mathbb{R}_+)^N: \sum_i h_i =1\}$. \\

Given an image and a skeletonization procedure, different spatial graphs can be built, depending on the weights assigned to the edges and the measure assigned to the nodes. In Figure \ref{fig:skeletonization}, we show an example of fibrotic patch turned into a simplified skeleton (obtained by removing all the nodes of degree exactly $2$). Given the spatial graph, we assign as the edge attribute a linear approximation of the area under the edge $(l_{lk}*w_{lk})$ divided by the total area of the patch $a$: $w_{lk}=(l_{lk}*w_{lk})/a, \forall l,k =1, \dots,N$ where $N$ is the number of nodes. The measure $\mu$ on the nodes is uniform with weight $1/N$. The function $d_X$ is the weighted shortest path distance between each couple of nodes. The weighted shortest path distance is one of the most common distances defined on a graph and it is computed as the length of the weighted shortest paths between every pair of nodes.

\subsection{Fused Gromov-Wasserstein Distance}
Spatial graphs have nodes attributes in $\mathbb{R}^p$ ($p=2$ in our case), edges attributes in $\mathbb{R}_{+}$, and they have different number of nodes and edges. A candidate metric between two spatial graphs should be able to compare both the spatial attributes, the edge attributes, and the graph structures. For such purpose, we adapt the Fused Gromov Wasserstein metric to spatial networks (introduced in a more general setting in \cite{vayer2019optimal,vayer2020fused}). This distance merges a spatial distance between the node attributes and a distance between the edges optimizing over possible coupling of the nodes.
\begin{definition}\label{def:fgw}(Fused Gromov-Wasserstein Distance for Spatial Graphs) Given two graphs $(X\times A, d_X, \mu)$ and $(Y\times B, d_Y, \nu)$, the Fused Gromov-Wasserstein (FGW) distance between spatial graphs is defined as:
\begin{equation}\label{eq:fgw}
    d(\mu,\nu)=\min_{\pi \in \Pi(\mu,\nu)} E(\pi)
\end{equation}
    where:
    $$E(\pi)=\sum_{\substack{i,k\in X \\ j,l \in Y}}[(a_i-b_j)^2 +(d_X(i,k)-d_Y(j,l))^2] \pi_{ij}\pi_{kl}$$
    In details,
    \begin{itemize}
        \item[-] $(d_X(i,k)-d_Y(j,l))^2$ accounts for the distance between the shortest paths from $i$ to $k$ in graph $X$ and from $j$ to $l$ in graph $Y$;
        \item[-] $(a_i-b_j)^2$ is the distance between the spatial coordinates of the nodes $i,j$ - in our case the Frobenious norm;
        \item[-] $\pi_{ij}\pi_{kl}$ is a joint distributions over $(X\times A) \times (Y\times B)$, assigning a probability over the node couplings (i in X coupled with j in Y; k in X coupled with l in Y ).
    \end{itemize}
\end{definition}
The definition is following \cite{vayer2020fused}, adapted to the discrete case here considered. The metric is null if two graphs have the same number of nodes, uniform weights on the nodes, and a one-to-one mapping between the nodes of the graphs which respects both the shortest paths and the spatial attributes. As proven in \cite{vayer2020fused}, the distance converges to a local minimum in the discrete case. 
\begin{remark}
    The FGW distance satisfies the triangular inequality up to a $2$ factor (See Proposition 5.3 in \cite{vayer2020fused})
\end{remark}

\subsection{Invariance under Rotation Actions}
The development of the fibrotic tissue depends on the orientation of the muscle fibers \citep{okenov2025analysis}. Given two spatial graphs, we would like to compare them up to rotation. While the Gromov Wasserstein distance is invariant under rotation (see \cite{memoli2011gromov} for details about GW and its application to object matching), the FGW distance considers the spatial attributes as part of the distance between two objects. The resulting FGW metric is not invariant under rotation. To overcome this limitation, we propose an alignment procedure of the spatial graphs. An orthogonal matrix $O\in \mathcal{O}(2)$ can be applied to the set of nodes $A\in \mathbb{R}^{N \times 2}$ via multiplication $AO^T$, resulting in a set of rotated coordinates  $A^r\in \mathbb{R}^{N \times 2}$ (and consequently a rotated graph). Given this pre-alignment, we can redefine the distance as follows:

\begin{definition}[Context Informed Rotated FGW] Consider two spatial graphs $(X\times A, d_X, \mu)$ and $(Y\times B, d_Y, \nu)$. Given two orthogonal matrices $O_A, O_B\in O(2)$ representing the principal axes of variation of the two sets of spatial nodes $A,B$, the context informed rotated FGW distance  is equal to $$d(\mu,\nu)=\min_{\pi \in \Pi(\mu,\nu)} E(\pi)$$
    where:
    $$E(\pi)=\sum_{i,j,k,l}  [([AO_{A}^{T}]_i-[BO_B^T]_j)^2 + (d_X(i,k)-d_Y(j,l))^2]\pi_{i,j}\pi_{k,l}$$
    where $[AO_A^T]_i$  and $[BO_B^T]_j$ are the i-th and the j-th elements of the rotated coordinate matrices.
\end{definition}

Note that we have a different $O\in \mathcal{O}(2)$ per each graph: there is no reciprocal matching between spatial networks but a context informed alignment applied to each graph separately. Given a graph, we run the following algorithm before performing any analysis:
\begin{algorithm}
\caption{Context Informed Alignment Algorithm}\label{alg:fgw_withalignment}
\begin{algorithmic}
\Require $(X\times A, d_X, \mu)$ 
\State Find the principal axis of the spatial coordinates via $A = USV^T$
\State Compute the rotated spatial coordinates as $A^r=AV$
\State Update the data representation as $(X\times A^r, d_X, \mu)$ 
\end{algorithmic}
\end{algorithm}

\begin{figure}[ht]
\begin{subfigure}{.45\linewidth}
  \centering
  \includegraphics[width=.99\linewidth]{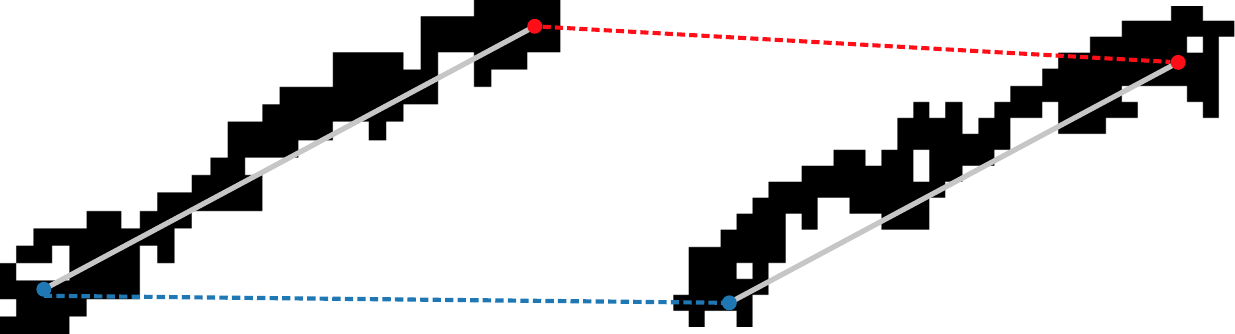}
        \vspace{1cm}
  \caption{Most similar graph in the set}
\end{subfigure}
\begin{subfigure}{.45\linewidth}
  \centering
  \includegraphics[width=.99\linewidth]{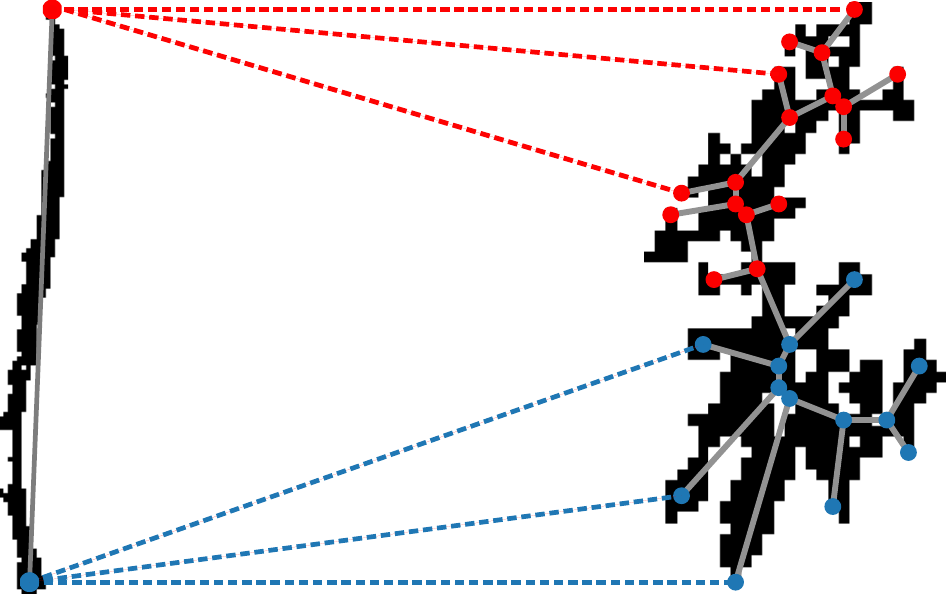} 
  \caption{Most dissimilar subgraph in the set}
\end{subfigure}

\caption{Example of the two most similar (left) and the two most dissimilar (right) graphs. The color of the nodes encodes the corresponding node matching proposed by the Fused Gromov-Wasserstein distance after aligning the graph with respect to their principal axes.}
\label{fig:example_distances}
\end{figure}

\section{Metric statistics for spatial graphs}
\label{sec:metric_statistics}
As detailed in the introduction, cardiac fibrosis is mostly unexplored in terms of its geometry and topology within human heart. Thanks to the metric defined in Section \ref{sec:spatial_graphs}, we have a metric function to compare the set of spatial graphs representing fibrotic tissue. Firstly, we want to test if there are differences across patients. Then, we can detail the analysis by exploring the ventricle spatially: a vertical comparison consists in comparing the spatial graphs in the different slices; an horizontal comparison consists in comparing the spatial graphs according to their distance from the boundary of the ventricle. To run all these comparisons between groups, we perform an ANOVA procedure for spatial graphs. Secondly, we look at the data within each group to understand where the differences are. As the spatial graphs are a simplified model of the original fibrotic patches, we would like to compare the groups by looking at the images directly. Instead of computing a centrality estimator such as the Fréchet Mean, we use a depth measure to identify multiple representatives of the empirical distribution. Such representatives mimic the sample quantiles in the Euclidean setting and correspond to actual spatial graphs observed in the set. To summarize the empirical distribution, we introduce the DepthPlot. The DepthPlot mimic a box-plot in the univariate setting, providing an intuitive visualization of the depth measure for data in metric spaces. 

\subsection{ANOVA for Spatial Graphs}
Given a metric space, we are interested in testing differences between groups. We opt for a generalization of the classical analysis of variance to metric spaces: see for example \cite{huckemann2009intrinsic} for shape data, \cite{cuevas2004anova} for functional data, and \cite{ginestet2017hypothesis} for graph data. Following the latest work of \cite{mueller2024anova}, we can use Anderson's  ANOVA for spatial graphs equipped with FGW metric. Anderson's ANOVA is a more proper way of referring to PERMANOVA (i.e. Permutation Multivariate ANOVA \citep{Anderson2014permutational}) in general metric spaces, where the multivariate adjective is not appropriate. The idea behind Anderson's ANOVA is to extend the standard Euclidean ANOVA to general pairwise dissimilarities. Note that the Anderson's ANOVA only relies on a metric function, making this method applicable in our setting.\\

Consider a set of $n$ spatial graphs $S_1,\dots,S_n$ in Spatial Graph Space $(\mathcal{S},d)$ - where $S_i=(X_i\times A_i,d_{X_i},\mu_i)$; consider $d=[d_{ij}]$ the distance matrix, where the $i,j=1,\dots,n$ element is the context informed rotated FGW distance between the spatial graphs. Given a partition  $\Pi_1, \dots, \Pi_k$ in $K$ groups, we would like to test the following hypothesis: $$ H_o: P_{\Pi_l}=P_{\Pi_m} \quad \text{vs} \quad H_1: P_{\Pi_l}\neq P_{\Pi_m}$$
where $P_{\Pi_l}$ and $P_{\Pi_m}$ with $l,m =1,\dots,K$ are $K$ potentially different distributions of spatial graphs (for example the partition into patients). We can build a test based on FGW distance matrix by computing the following test statistics:
$$F=\frac{n-k}{k-1}\frac{SS_A}{SS_T}$$
 $SS_T=\frac{1}{n}\sum_{i=1}^{n}\sum_{j=i+1}^{n}d_{ij}$ represents the total sum of distances. $SS_A=SS_T-SS_W$ represent the sum of square distances across the groups, which can be computed via the sum of distances within the groups $SS_W=\sum_{l=1}^{k}\frac{1}{2*|\Pi_l|}\sum_{i,j\in \Pi_l}d_{ij}$. The p-value is computed using the permutation testing procedure (see \cite{pesarin2010permutation} for a full introduction to the topic). The Anderson's ANOVA allows for a comparison across patients as well as a spatial comparison across slices and within slices, detailed in Section \ref{sec:results}.

Before we proceed in defining the sample depth measure for spatial graphs, an important remark is needed. In generic metric spaces, a large corpus of papers bases the test statistics on comparing the distributions directly (e.g. test based on the distance profiles by \cite{dubey2024metric}; test based on energy functions by \cite{gretton2012kernel,lovato2021multiscale}). Unfortunately, these tools rely on the metric to be of negative type (see \citep{lyons2013distance} for an introduction to the concept). 
\begin{proposition}\label{prop_fgw}
The Fused Gromov Wasserstein metric is not of negative type.
\end{proposition}
To prove the metric to be of negative type, both the two components have to be of negative type. The first component - i.e. the distance between the spatial coordinates on the nodes - is not of negative type as for Proposition 8.2 in \cite{peyre2019computational}. As a consequence, the FGW metric for spatial graphs is not of negative type, making all these test not applicable in this context.

\subsection{DepthPlot for Spatial Graphs}
Once the testing procedure is performed, we look at the set of spatial graphs to actually understand and spot the differences between the empirical distributions. Data depth is a powerful exploratory tool which allows to rank the observation based on how ``peripheral'' or ``central'' a point is within a point cloud. The higher the depth the more representative the point is of the distribution - recalling a median point in the Euclidean setting. Similarly, a low depth allows to identify points that are assimilated to "outliers". For non-Euclidean data, different concept of depth have been introduced in the literature: for functional data \citep{fraiman2001trimmed,lopez2009concept}, directional data \citep{pandolfo2018distance}, positive definite matrices \citep{chau2019intrinsic}. For data on metric spaces, such a spatial graphs, there are very few depth measures available: Metric Half Space (MHS) depth \citep{geenens2023statistical,dai2023tukey} and spatial depth \citep{virta2023spatial}. Due to its impact in the multivariate setting, we opt here for adapting the MHS depth to spatial graphs (please see the recent contribution of \cite{dai2023tukey} for an overview of MHS depth for object data). 

\begin{definition}
    Given a set of $n$ spatial graphs $S_1,\dots,S_n$ in Spatial Graph Space $(\mathcal{S},d)$, the sample metric depth at $S\in \mathcal{S}$ is $$D_n(S)=\inf_{S_1,S_2\in \mathcal{S}, d(S_1,S)\leq d(S_2,S)} \frac{1}{n}\sum_{i=1}^{n}I\{d(S_i,S_1)\leq d(S_i,S_2)\}$$
\end{definition}
The depth measure assign a value to each graph in the sample. The higher the value the more central the graph is with respect to the empirical distribution of the spatial graphs.
\begin{figure}[htbp!]

\begin{subfigure}{.6\linewidth}
\centering
  \includegraphics[width=\linewidth]{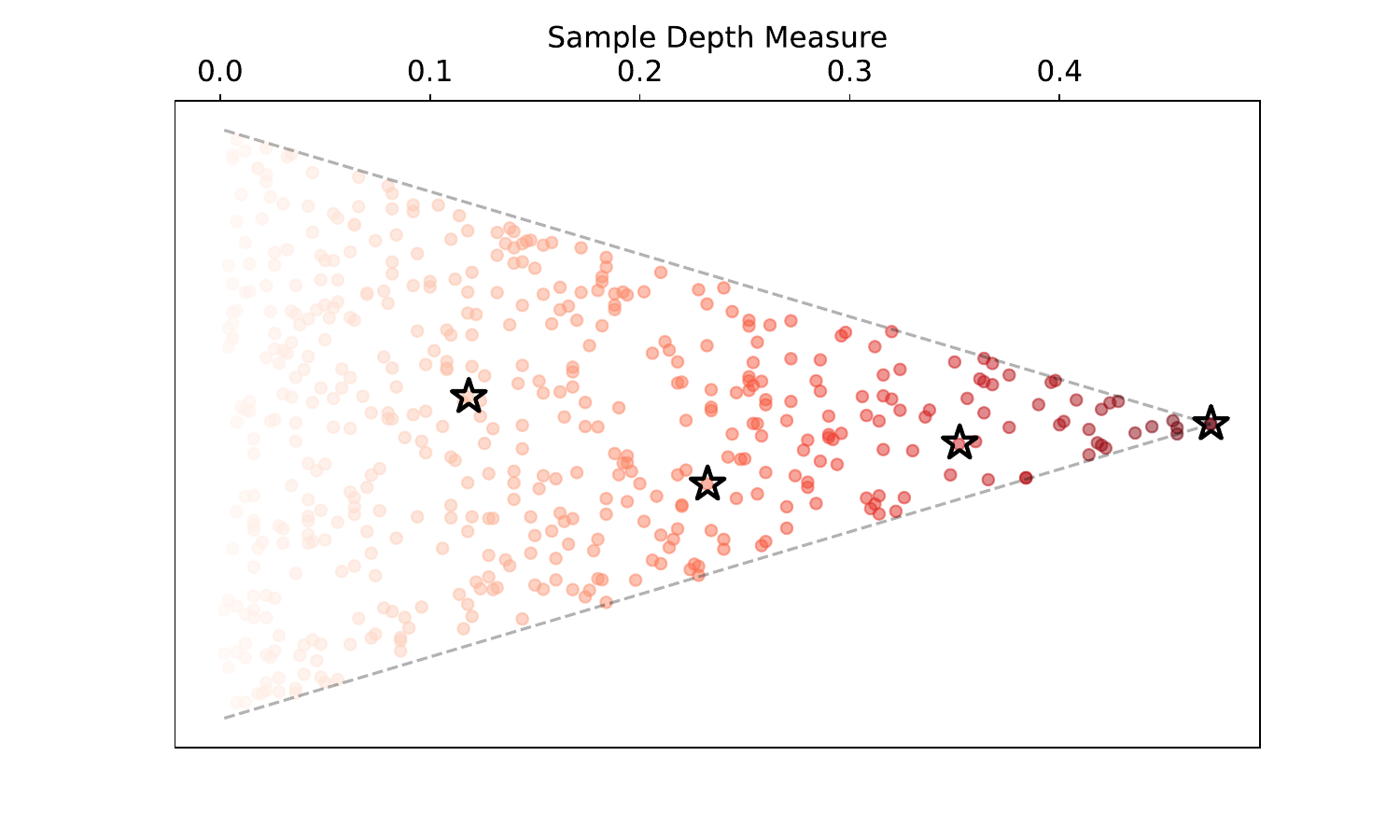}
\end{subfigure}%
\begin{subfigure}{.4\linewidth}
\hspace*{-0.6in}
\centering
  \includegraphics[width=.9\linewidth]{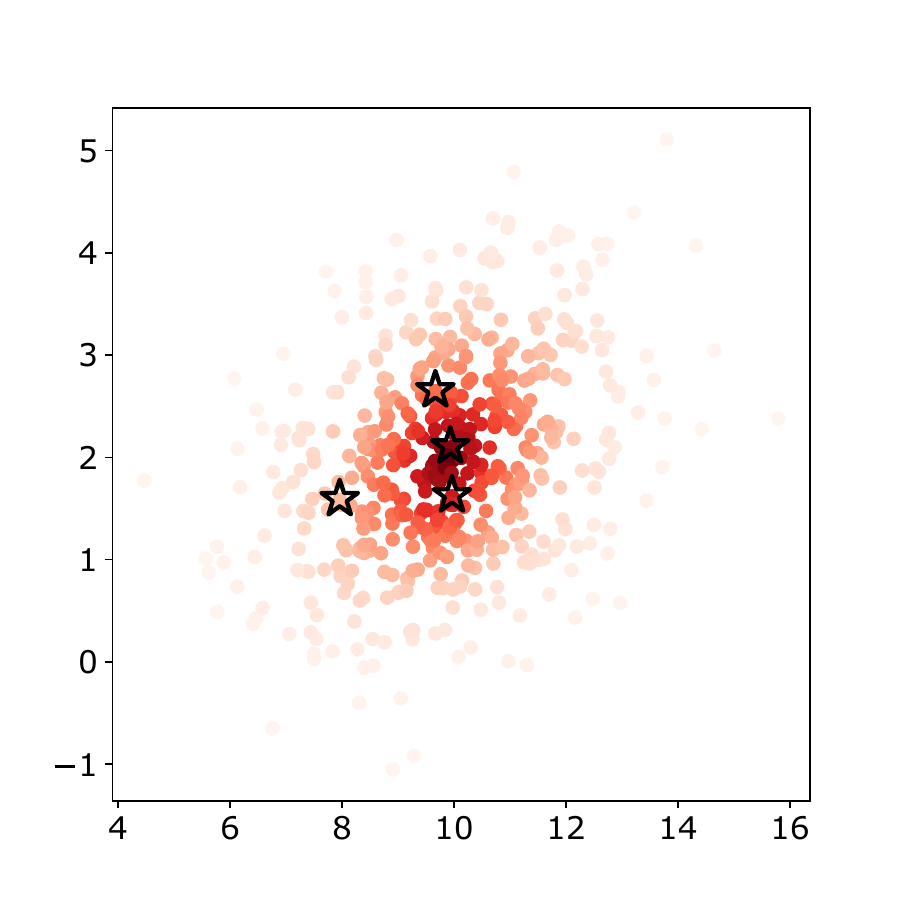}
\end{subfigure}
\caption{DepthPlot: an example of sample depth values of $500$ points from a bi-dimensional Normal distribution. Left: the sample depth values are visualized by jittering the points along the x-axis. The cone shape alludes to the depth. Right: the scatterplot of the points in $\mathbb{R}^2$, colored by their corresponding depth values. The stars are representative points (min, $0.25$, $0.75$, max depth values).}
\label{fig:DepthPlot}
\end{figure}
The MHS Depth has been already applied to non-Euclidean data such as trees embedded in BHV space \citep{dai2023tukey}. However, there is a lack of visualization tools. We introduce DepthPlot, a visualization tool for non-Euclidean data (see Figure \ref{fig:DepthPlot}). Given a set of objects - in our setting a set of spatial graphs - and the MHS depth, the DepthPlot allows to show the depth and the amount of points per depth value via a scatterplot of jitter points on a cone structure. In Figure \ref{fig:DepthPlot}, we design a simple experiment to show how to build a DepthPlot. We sample $500$ points from $\mathcal{N}_2(\mu,\Sigma)$ where $\mu=(10,2)$ and the covariance matrix is $\Sigma=((3,0.5),(0.5,1))$. Then, we compute the MHS depth. In the right panel of Figure \ref{fig:DepthPlot}, the cone structure alludes to the depth, positioning deeper points at the tip of the cone and marginal points at the base of the cone. Along with this first scatterplot, a set of $M$ objects - in this example a set of $M$ points in $\mathbb{R}^2$ - are selected to visualize representatives of different depth values. The $M$ representative are visualized as stars in the DepthPlot and can be selected in different ways (e.g using the depth quantiles, clustering the depth values). For this explanatory example, we can visualize the points cloud in $\mathbb{R}^2$ colored according to the MHS depth highlight the representative points in star shapes. The depth plot can be used for every data object equipped with a metric (e.g. images, trees, and graphs). If an embedding in $\mathbb{R}^2$ is not available, the representative points can be visualized directly. The DepthPlot is used in Section \ref{sec:results} for the interpretation of the test outcome via visualization of the sample distributions of spatial graphs.

\section{Results on Cardiac Fibrosis Shape}
\label{sec:results}

\begin{figure}[htbp!]
    \centering
    \includegraphics[width=0.5\linewidth]{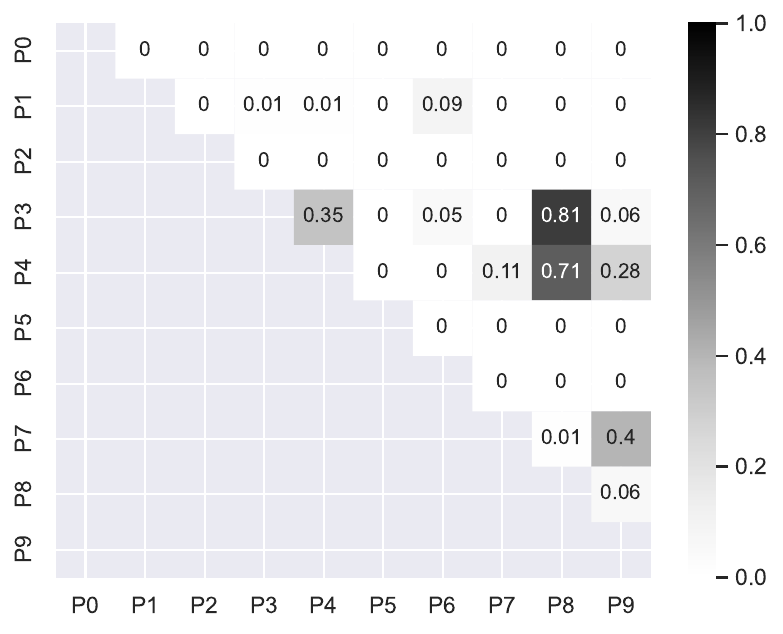}
    \caption{Test across patients: every row and column represent a different patient. We report the p-values of the Anderson's ANOVA two-sample test across the patients.
    }
    \label{fig:patient_test}
\end{figure}

We consider a subsample of $n=15684$ spatial graphs $\{S_1, \dots, S_n\}$, obtained by selecting only up to $100$ fibrotic patches per slice with an area between $100$ and $500$ pixels.  The software implementation is available in the GitHub repository spatial\_graph \citep{calissano2026git}. As detailed in Section \ref{sec:data}, the objective of the analysis is to compare the set of patients to understand how these structures differs in terms of geometry and topology. All the patients have been selected as they showcase specific genetic mutation, so clinicians expect to see difference across the patients in the way the cardiac fibrosis develops. We start by computing the distance matrix using the Fused Gromov-Wasserstein distance with the context informed alignment algorithm (Algorithm \ref{alg:fgw_withalignment}). Once the distance is computed, we compare the spatial graphs across the $10$ different patients. In Figure \ref{fig:patient_test}, we report the p-values of the  Anderson's ANOVA test ($500$ permutation runs per test). The null hypothesis is rejected in almost all cases.\\

\begin{figure}[htbp!]
     \centering
     \begin{subfigure}{\linewidth}
         \centering
         \includegraphics[width=\linewidth]{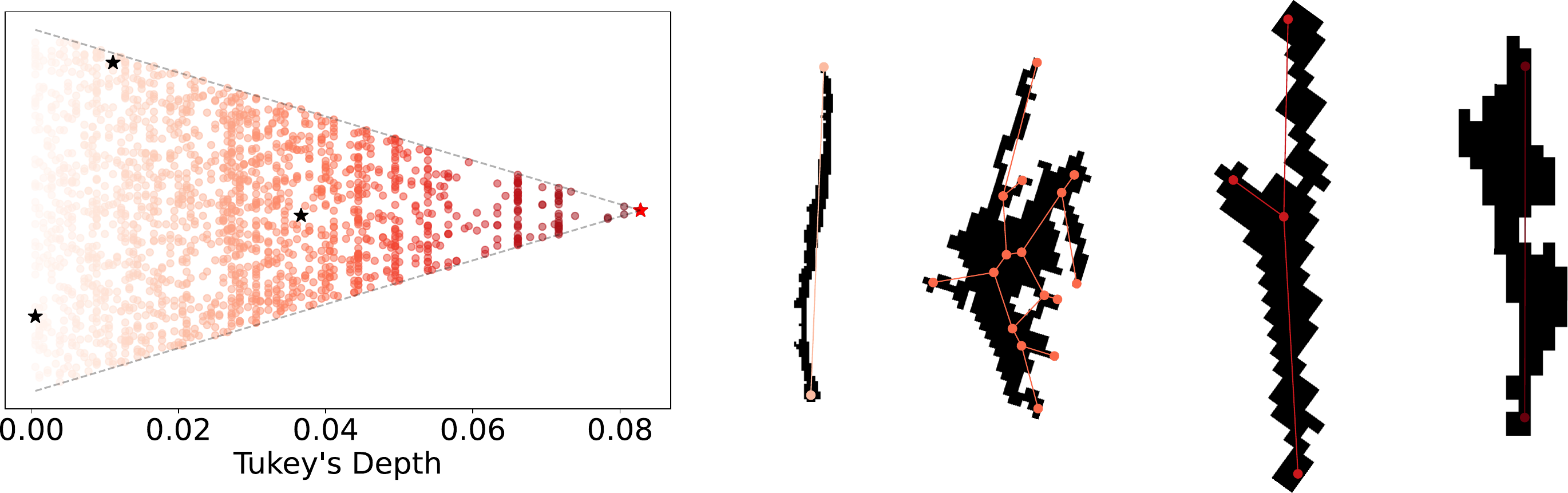}
         \caption{DepthPlot of Patient 2.}
         \label{fig:DepthPlot_P2}
     \end{subfigure}
     \begin{subfigure}{\linewidth}
         \centering
         \includegraphics[width=\linewidth]{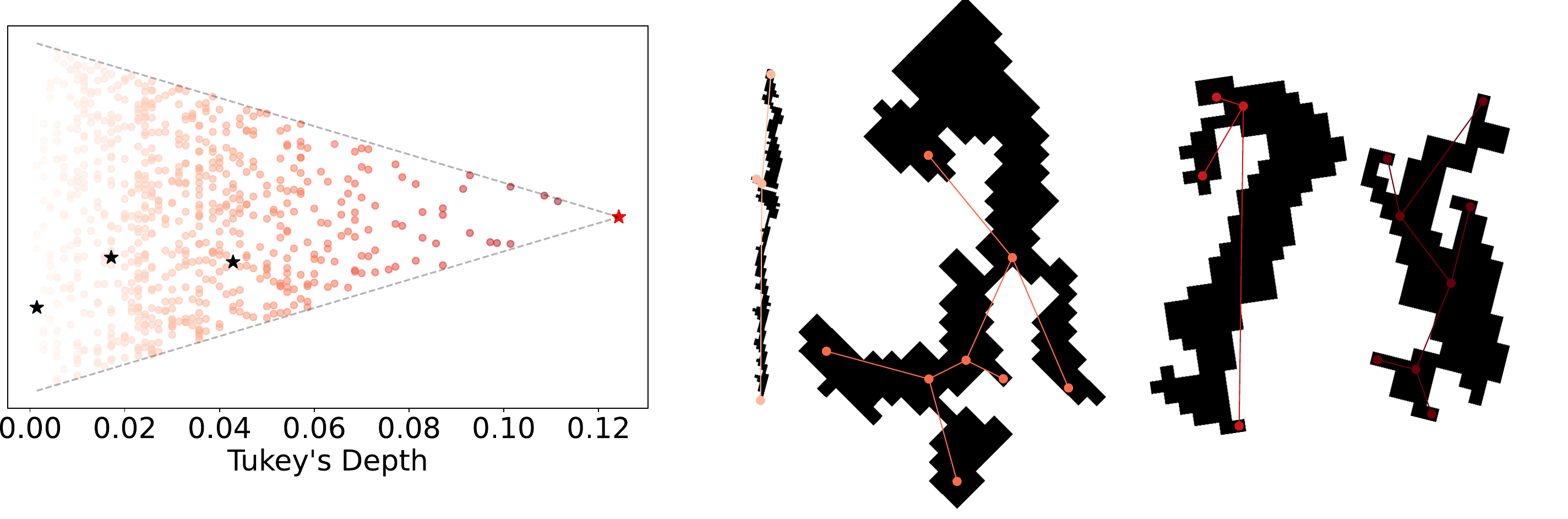}
         \caption{DepthPlot of Patient 3.}
         \label{fig:DepthPlot_P3}
     \end{subfigure}
          \begin{subfigure}{\linewidth}
         \centering
         \includegraphics[width=\linewidth]{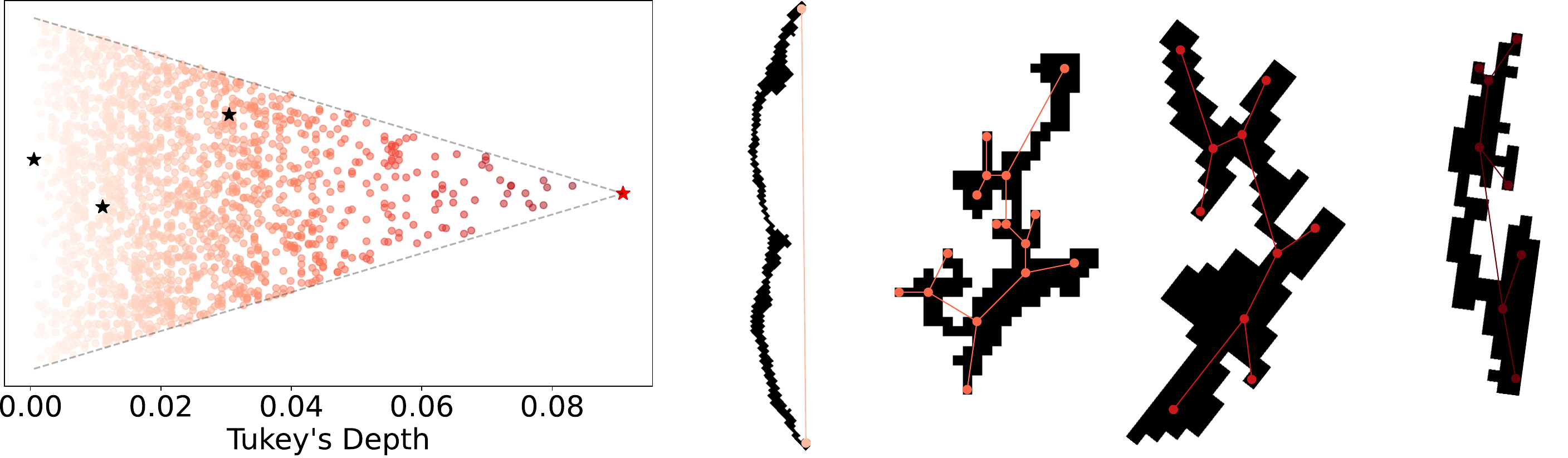}
         \caption{DepthPlot of Patient 8.}
         \label{fig:DepthPlot_P8}
     \end{subfigure}
    \caption{DepthPlot of Patient 2, Patient 3, and Patient 8 using MHS depth. The spatial graphs showed on the left correspond to: the  minimum sample depth value, the $0.25$ quantile of the depth values, the $0.75$ quantile of the depth values, the maximum sample depth value (i.e. the median spatial graph).}
    \label{fig:DepthPlot_plots}
\end{figure}

As the test outcome via the p-value does not provide any tangible interpretation of the difference across the sample distributions, we can use MHS depth measure and the DepthPlot to visualize the distribution of the spatial graphs within each patient. We compute the MHS depth for each patient: the result is a positive value associated to each spatial graph representing its depth within the sample distribution. In Figures \ref{fig:DepthPlot_P2}, \ref{fig:DepthPlot_P3} and \ref{fig:DepthPlot_P8}, we show the distribution of the spatial graphs in Patient 2, Patient 3, and Patient 8, along with $M=4$ representative spatial graphs. The representative spatial graphs correspond to: the minimum depth value, the $0.25,0.75$ depth quantiles, and the maximum depth value (i.e., the median spatial graph).\\

Patient 3 and Patient 8 has the same genetic mutation (LMNA) while Patient 2 has a different one (RBM20). Both LMNA and RBM20 are well-established genes in which pathogenic variants are linked with a strong fibrotic component (see for example \cite{verdonschot2025clinical} for a recent study). We select such patients as representatives to interpret the test results. In the Anderson's ANOVA testing, the test between the sample distribution in Patient 2 and Patient 3 has very low p-value, showing evidence to reject the null hypothesis, while the test between the sample distribution of Patient 3 and Patient 8 has a very high p-value, showing not enough evidence to reject the null hypothesis. By looking at the three DepthPlot, we can see that all the three sample distributions share similar peripheral spatial graphs: the minumum depth graphs are all elongated thin shapes and the $0.25$ depth quantile plot are highly fragmented shapes. However, the most representative spatial graphs of the three distributions are the last two (i.e. the spatial graphs corresponding to the $0.75$ depth quantile and the maximum depth value). In Patient 3 and Patient 8, the last two plots share similar structures, with profiles that are more fragmented than the one shown in Patient 2. In fact, Patient 2 has a sample distribution centered around simple shape resulting into a one edge spatial graph while Patient 3 and Patient 8 have a median spatial graph with nodes of degree 2 and 8 nodes. 

\subsection{Vertical and Horizontal Spatial Exploration}
The spatial graphs have themselves a position within the slices and within the ventricles. Such information is  important in the analysis of the fibrotic textures. In fact, orientation of myocardial fiber at given point is determined by the position of the point  with respect to the outer and inner surface of the heart. In turn, the local orientation of myocardial fiber can have  impact on  the shape of the cardiac fibrosis, as  it can initially develop along the fiber direction. To study the effect of the spatial position within the heart, we perform testing across the slices on three representative patients (Patient 2,3, and 8).\\

\begin{figure}[htbp!]
\begin{subfigure}[b]{0.30\linewidth}
\includegraphics[width=\linewidth,trim={0 0 6cm 0},clip]{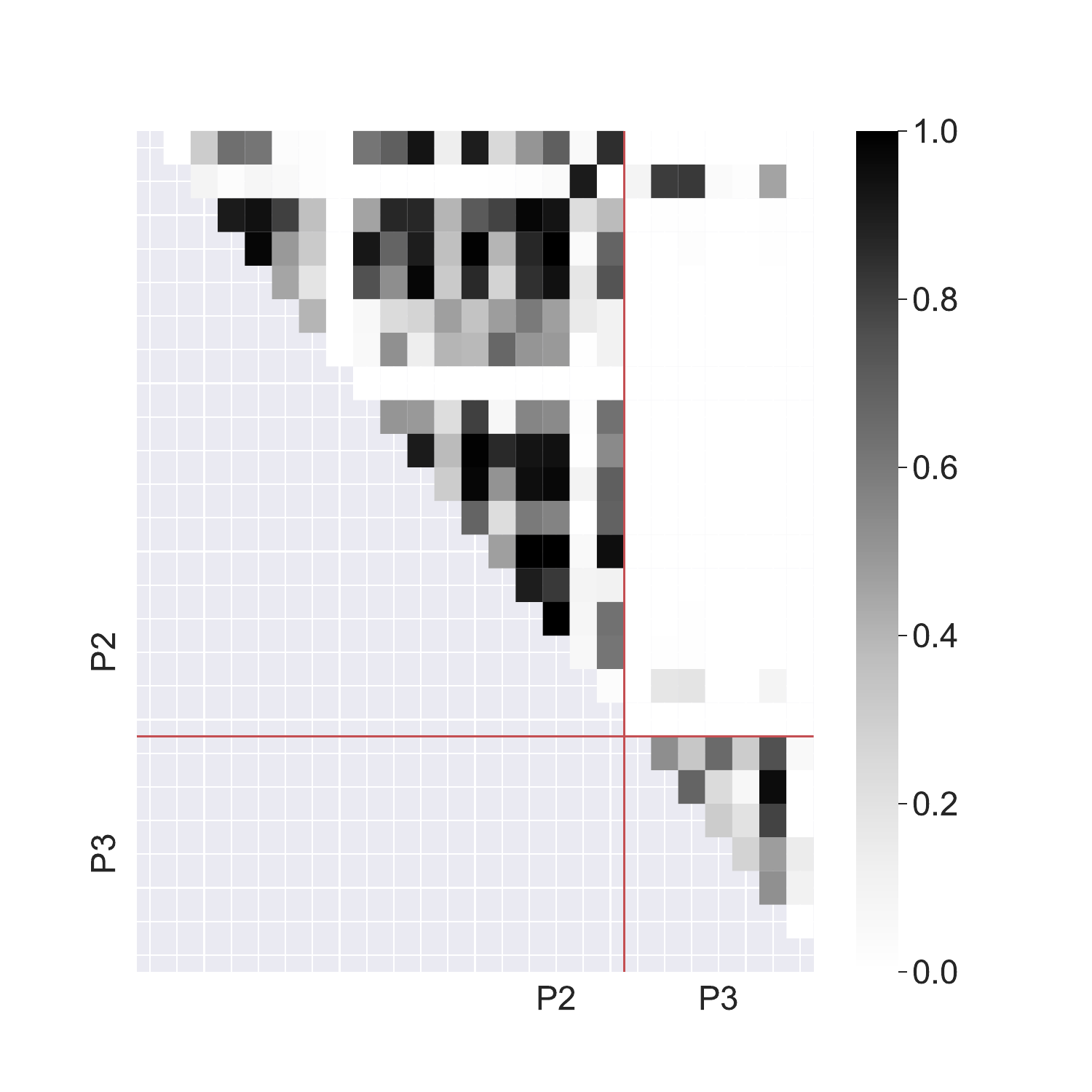}
\end{subfigure}
\begin{subfigure}[b]{0.30\linewidth}
\includegraphics[width=\linewidth,trim={0 0 6cm 0},clip]{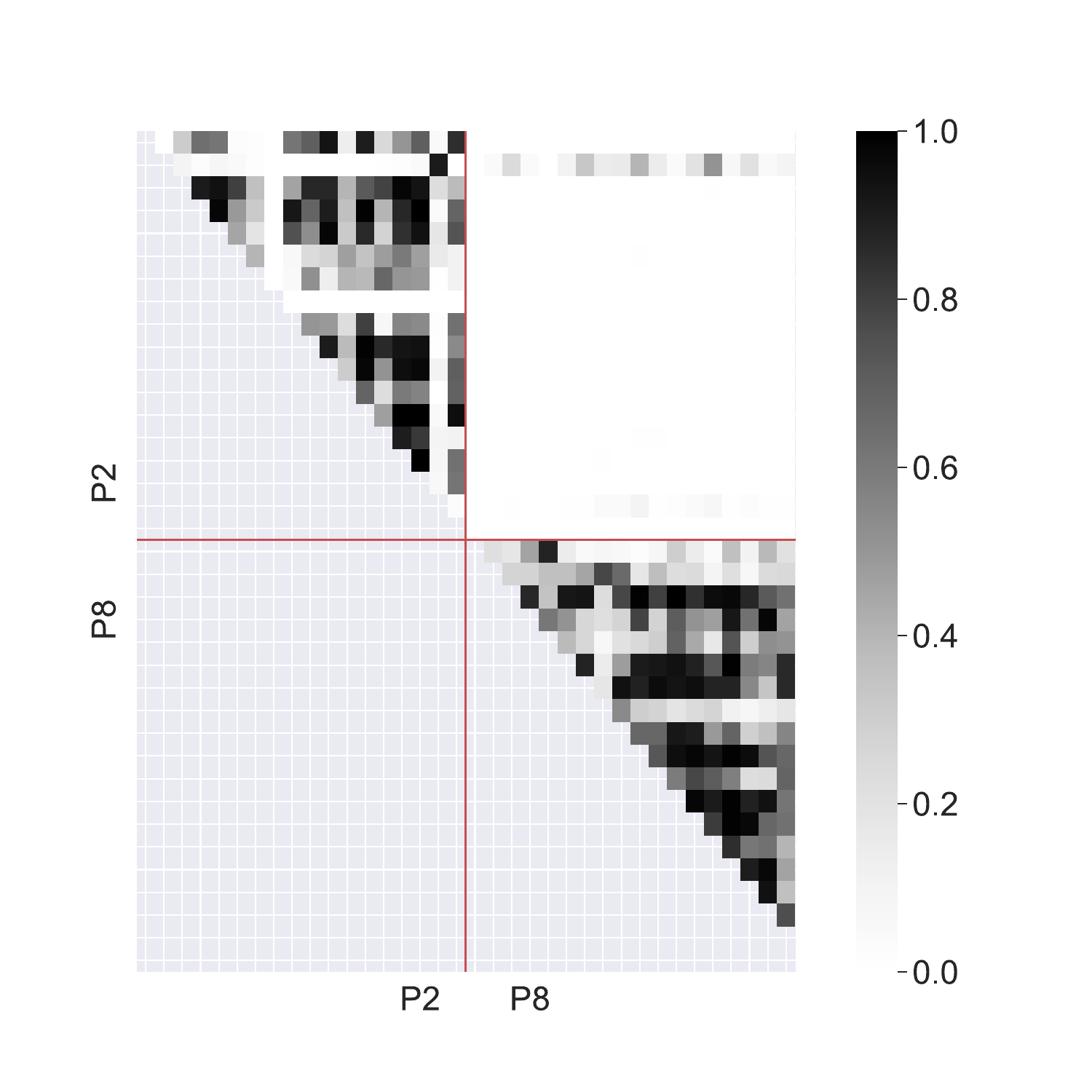}
\end{subfigure}
\begin{subfigure}[b]{0.30\linewidth}
\includegraphics[width=\linewidth,trim={0 0 6cm 0},clip]{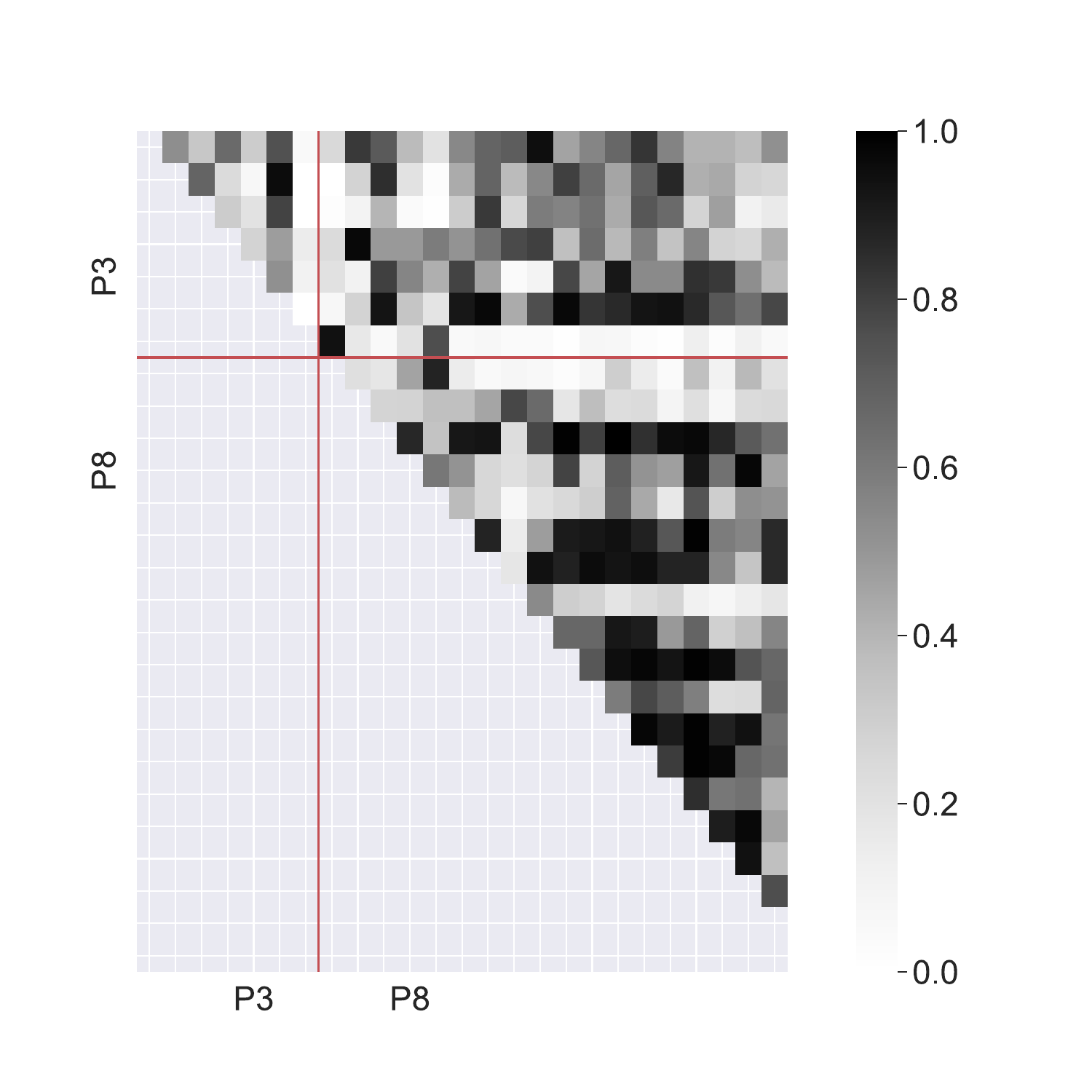}
\end{subfigure}
    \caption{Vertical Exploration: Anderson's ANOVA of spatial graphs across the different slices. We report the p-values of the test across each slice across three selected patients (Patient 2, Patient 3, and Patient 8).}
    \label{fig:test_slices}
\end{figure}

In Figure \ref{fig:test_slices}, we display the Anderson's ANOVA p-values of all the tests across all the slices of the three selected patients. The test is addressing the question of the spatial graphs coming or not from the same distribution across the different slices. The number of slices are different in every patients because the hearts have different sizes.
The first two figures shows that if the null hypothesis is rejected in the global test between patients, it is also reject in almost all the slices (see for example Patient 2 and Patient 3). Similarly, if there is no evidence for rejecting the null hypothesis in the Patient 3 vs Patient 8 test, the same result is observed in all the tests across slices.\\

\begin{figure}[htbp!]
    \centering
    \includegraphics[width=\linewidth]{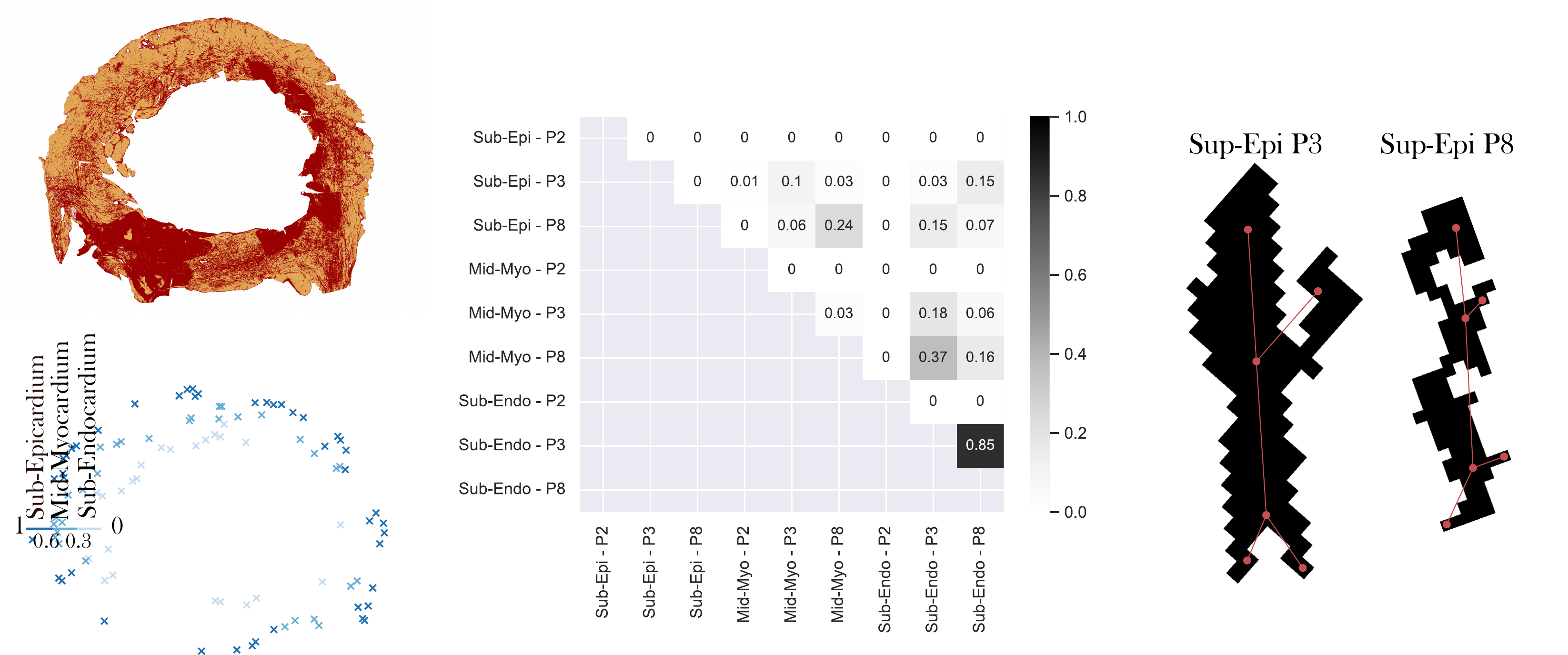}
    \caption{Horizontal Exploration: Anderson's ANOVA of spatial graphs across the Sub-Epicardium, Mid-Myocardium, and Sub-Endocardium. Left: We highlight the different area on a representative slide. Middle: The p-values of the test across each of the horizontal surfaces across the three selected patients (Patient 2, Patient 3, and Patient 8). Right: The median spatial graph of the Sub-Epicardium in P3 and P8, where the null hypothesis is rejected.}
    \label{fig:test_distance}
\end{figure}

These test across the slices allows for a vertical comparison. When working within a slice, an important parameter in the literature is the distance from the outer heart surface, ranging from 0 (outer surface) to 1 (inner surface). Such parameter divides the ventricle into three regions \citep{glashan2018whole}: Sub-Endocardium, Mid-Myocardium, and Sub-Epicardium (the internal part, the middle and the external part of what is called Myocardium). We start by performing a test, dividing the spatial graphs of Patient 2, Patient 3 and Patient 8 into three groups according to their position within the slices (Sub-Endocardium $[0,0.3)$, Mid-Myocardium $[0.3,0.6)$, and Sub-Epicardium $[0.6,1]$). By looking at the p-values in Figure \ref{fig:test_distance}, we null hypothesis to be rejected for all of the three areas in Patient 2, showing a consistency in the horizontal and vertical exploration. The null hypothesis is also rejected in the Sub-Epicardium in Patient 3 and Patient 8, while there is not statistical evidence to say Patient 3 and Patient 8 are different in the Mid-Myocardium and the Sub-Endocardium.

To summarize, the cardiac fibrosis studied as isolated fibrotic tissues in 2-dimensional slices is characterized by high variability and differences across patients, across slices, and within slices. We detail the analysis for two patients with the same LMNA genetic mutation, namely Patient 3 and Patient 8, and Patient 2 with RBM20 mutation. The vertical exploration support the hypothesis that there is a vertical difference across all the slices between Patient 2 and the other patients. The horizontal exploration underline a difference across all the three surfaces between Patient 2 and the other patients while Patient 3 and Patient 8 show differences only in the Sub-Epicardium.

\section{Comparison with Topological Features Extraction}
\label{sec:comparison}
\begin{figure}[htbp!]
    \centering
    \includegraphics[width=0.5\linewidth]{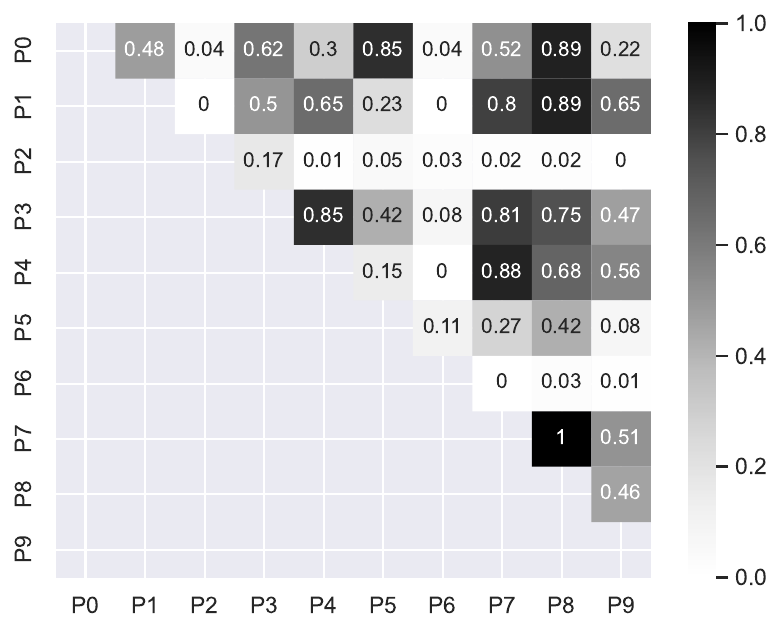}
    \caption{Test across patients based on topological and geometrical feature extraction: every row and column represent a patient. We report the p-values of the Anderson's ANOVA two-sample test across patients.}
    \label{fig:p_val_topo}
\end{figure}

The choice of modeling porous, irregular, and complex tissue structures as spatial graphs is not unique. Another solution - which is apparently more intuitive - is to summarize the images using a set of topological and geometrical features. In the following paragraph, we compare the testing results with another data representation. Starting from the exact same set of images $I_i\in \mathbb{R}^{n\times m}, \quad i =1,\dots,15684$, described in Section \ref{sec:data}, the fibrotic tissue is now represented as a vector containing topological and geometrical characteristics. In particular, we estimated the area, the perimeter, and the Euler characteristic. Such three features arise naturally as the Minkowski functionals for a shape embedded in $\mathbb{R}^2$ \citep{mecke2000additivity}. We perform the same ANOVA testing procedure described in the previous section based on the Euclidean distance between these features. The resulting p-value map is not able to underline the expected differences between patients with different genetic mutation, as shown in Figure \ref{fig:p_val_topo}. Despite the complexity in modeling the tissue as spatial graph, the simple feature extraction procedure is not able to capture the intrinsic and expected differences between the patients. 

\section{Discussion}
\label{sec:conclusion}
We introduced a metric space for the analysis of spatial graphs — graphs with nodes embedded in $\mathbb{R}^2$ or $\mathbb{R}^3$. Such graphs can be used as a simple model to summarize the geometry and topology of a shape, which displays a fragmented and porous structure, hard to analyze with standard shape analysis tools. We used spatial graphs to study the shape of cardiac fibrotic tissue within the human heart, basing our analysis on a unique dataset of histopathological images of human hearts. Given the introduced metric space, we performed a two-sample Anderson's ANOVA to compare different patients as well as different areas of the heart. To interpret the two sample test outcome, we used data depth exploring the empirical distribution of the spatial graphs. We also defined DepthPlot as an intuitive visualization tool of the depth measure. Our results showed that the testing procedure identified differences across subjects, reflecting certain genetic mutations, and the DepthPlot allows the identification of the most representative spatial graphs per patient, per slice, and per area of the ventricle.\\

Beyond the present study, the metric space introduced here may be applied to compare other tubular or complex porous structures, making the spatial graph space a highly adaptable framework with potential applications across diverse domains. From a translational perspective, this study provides the first quantitative characterization of the fine geometry of cardiac fibrosis and lays the groundwork for future investigations. For instance, computational models of cardiac dynamics can be improved by incorporating realistic representations of fibrotic tissue. Furthermore, the proposed analysis can be extended to much smaller tissue specimens, including biopsy samples.

\section*{Acknowledgment}
We would like to thank Prof. Mikhael Gromov, Dr. Titouan Vayer, Dr. Yvo Pokern, and Prof. Karthik Bharath, for the fruitful discussions that helped me spotting the flows and the strengths in this work.

\end{document}